\documentclass[journal]{IEEEtran}

\usepackage{siunitx}

\ifCLASSINFOpdf
   \usepackage[pdftex]{graphicx}
   \graphicspath{{../Images/}}
   \DeclareGraphicsExtensions{.pdf,.jpeg,.png}
\else
\fi

\usepackage{amsmath}
\usepackage{amssymb}
\usepackage[online]{threeparttable}

\usepackage{algorithm}
\usepackage[noend]{algpseudocode}

\usepackage{array}

\ifCLASSOPTIONcompsoc
 \usepackage[caption=false,font=normalsize,labelfont=sf,textfont=sf]{subfig}
\else
 \usepackage[caption=false,font=footnotesize]{subfig}
\fi

\usepackage{xcolor}
\usepackage{siunitx}
\usepackage{cleveref}
\usepackage{bm}

\newcommand{\be}{\begin{equation}}
\newcommand{\ee}{\end{equation}}

\DeclareMathOperator*{\argmin}{arg\,min}

\begin{document}
\title{Learned Spectral Computed Tomography}

\author{Dimitris~Kamilis,%
        ~Mario~Blatter,%
        ~and~Nick~Polydorides%
\thanks{D. Kamilis (d.kamilis@ed.ac.uk) and N. Polydorides are with the Institute for Digital Communications, School of Engineering, University of Edinburgh, UK.}%
\thanks{M. Blatter is with the Department of Electrical Engineering and Information Technology, ETH Zurich.}%
}

%\markboth{IEEE TRANSACTIONS ON MEDICAL IMAGING}%

\maketitle

\begin{abstract}
Spectral Photon-Counting Computed Tomography (SPCCT) is a promising technology that has shown a number of advantages over conventional X-ray Computed Tomography (CT) in the form of material separation, artefact removal and enhanced image quality. However, due to the increased complexity and non-linearity of the SPCCT governing equations, model-based reconstruction algorithms typically require handcrafted regularisation terms and meticulous tuning of hyperparameters making them impractical to calibrate in variable conditions. Additionally, they typically incur high computational costs and in cases of limited-angle data, their imaging capability deteriorates significantly. Recently, Deep Learning has proven to provide state-of-the-art reconstruction performance in medical imaging applications while circumventing most of these challenges. Inspired by these advances, we propose a Deep Learning imaging method for SPCCT that exploits the expressive power of Neural Networks while also incorporating model knowledge. The method takes the form of a two-step learned primal-dual algorithm that is trained using case-specific data. The proposed approach is characterised by fast reconstruction capability and high imaging performance, even in limited-data cases, while avoiding the hand-tuning that is required by other optimisation approaches. We demonstrate the performance of the method in terms of reconstructed images and quality metrics via numerical examples inspired by the application of cardiovascular imaging.
\end{abstract}

\IEEEpeerreviewmaketitle

\section{Introduction}
\IEEEPARstart{S}{pectral} Photon-Counting Computed Tomography (SPCCT) has recently gained attention in medical imaging \cite{McCollough2015-rq, Danad2015-wi, Willemink2018-ll} as it was shown to provide a number of advantages compared to conventional X-ray Computed Tomography (CT), including material separation, beam-hardening artefacts removal and enhanced image quality. An important example is its ability to distinguish between various calcification levels in atherosclerotic plaques, which informs the risk for coronary artery disease \cite{Cormode2010-bv, Boussel2014-af}. SPCCT is based on the development of photon-counting detectors such as Medipix3 \cite{Ballabriga2011-xz}, which allow for resolving the energies of detected photons using a number of energy bins. The underpinning mathematical model accounts for the spectral properties of the materials, the source's spectrum and the detector sensitivity, and has the form of a non-linear attenuation law. 

Reconstruction methods for SPCCT can be categorised into: (i) direct, one-step reconstruction of the attenuation coefficient or material volume/mass fractions \cite{Long2014-np}, (ii) separate tomographic reconstruction for each energy bin followed by a material decomposition in the image domain \cite{Mendonca2014-bk}, and (iii) two-step schemes, based on material decomposition in the projection domain, which is followed by separate tomographic imaging of the materials \cite{Ducros2017-xa}. Out of these three approaches, the first is the most general, at the cost of being the most computationally expensive, whereas the second assumes an approximate forward model which can cause artefacts and deterioration of the image quality. As a result, in this paper we opt for the third, two-step approach, which adheres to the non-linear physics and additionally provides computational efficiency. 

The first step involves an unmixing process to resolve each material's projections (sinograms) from the measured spectral data and takes the form of an ill-posed, non-linear and generally non-convex problem, which poses challenges in terms of stability and computational efficiency. The solution of this problem can be addressed via variational methods as in \cite{Ducros2017-xa} where a regularised Gauss-Newton method is proposed. However, this requires a careful tuning of the regularisation parameters, otherwise the method may fail to recover a satisfactory solution. A more robust alternative is examined in \cite{Abascal2018-nl} which takes into account the non-convexity of the problem and suggests the solution of a series of locally convex subproblems, at the cost of increased computational complexity. However, such conventional regularisation methods are based on handcrafted terms that aim to capture some generic prior feature of the solution, but are not customised to the intrinsic features the solution exhibits for a specific task and dataset.

The second step involves the separate tomographic reconstruction of each material's volume or mass fraction based on the `metadata' obtained from the first step. This can proceed using variational approaches based on e.g. sparsity, or alternatively, using Deep Learning-based, data-driven methods, which have shown to offer state-of-the-art performance \cite{Ravishankar2019-fh, Arridge2019-wq}. A subclass of such schemes is based on unrolling an iterative optimisation algorithm and representing the updates as layers of a Neural Network (NN) \cite{Arridge2019-wq}. This results into learned algorithms which are trained to perform with high accuracy for specific tasks and datasets. Additionally, these algorithms can incorporate model knowledge for increased efficiency and accuracy, especially when data are limited. Such a scheme was recently proposed for conventional CT in \cite{Adler2018-xy} based on primal and dual variable updates, and was shown to outperform other approaches such as Total Variation or Filtered Back-Projection followed by some form of denoising.

In this paper, we propose an extension of the learned primal-dual algorithm in \cite{Adler2018-xy} to the SPCCT case, which takes the form of a NN-based algorithm that is customised to the task of imaging from case-specific spectral data, meaning data corresponding to an area of interest such as the heart. The method consists of two learned primal-dual algorithms, one for each of the steps (unmixing and imaging), for which we show that enhanced performance is feasible when they are integrated and trained end-to-end. The proposed method offers fast reconstruction, doesn't require hand-crafted regularisation terms or tuning of optimisation parameters and is optimised for the imaging task with high performance even for media comprising several materials. 

In support of these claims, we present numerical results for a toy-case of random-ellipse phantoms consisting of 5 materials, as well as for a medical phantom generated with the 4D XCAT software \cite{Segars2010-wa}. In particular, for the medical case we focus on the problem of identification and imaging of a number of materials in the heart area, including calcified plaques in coronary arteries \cite{Osawa2016-oo}. The existence of such plaques is associated with increased Coronary Heart Disease (CHD), whose risk depends on their size, position and composition \cite{Williams2019-le}. We show via numerical examples of a simplified plaque phantom the potential of learned SPCCT to identify and image plaques in the heart area and thus to provide quantitative information for quantifying CHD risks.

\section{SPCCT model}
We assume that the SPCCT imaging system consists of a radiation source with normalised photon flux density $S(E)$ (in photons per unit area-time-energy) and a photon-counting detector that can resolve detected photon energies into $N_b$ energy bins, with $D_b(E)$ the detector sensitivity in each bin, and $E$ the energy level. Taking into account the energy dependence of the linear attenuation coefficient (LAC) $\mu(x, E)$, results into the attenuation law for the expected number of detected photons in each bin \cite{McCullough1975-xc}
\be\label{model}
\bar{y}_b(\mu) = y_0 \int_0^{E_\text{max}} S(E) D_b(E)\, \exp \Bigl \{-\int_L\mu(x,E) \,\mathrm{d}x \Bigr \} \,\mathrm{d}E,
\ee
where $y_0$ is the initial source intensity (photons per unit time-area), $E_\text{max}$ the maximum source energy and the line integral is taken along an arbitrary line $L$ that starts from the source and ends on the detector. Taking into account the Poisson statistics describing the process, results into a measurement model given by the Poisson random variable $y=[y_1,\dots,y_{N_b}]$ with $y_b \sim \mathrm{Pois}(\bar{y}_b)$.
Although in theory the direct recovery of $\mu(x,E)$ from $y$ is possible, such a task faces very high computational costs due to the complexity of discretising a 3D or 4D variable simultaneously in the spatial and energy domains. A way to surpass this problem is to seek a separable representation as a sum of $N$ terms
\be\label{musum}
\mu(x,E)=\sum_{k=1}^N \mu_k(E) q_k(x),
\ee
which reduces significantly the computational complexity and permits the splitting of the problem into two steps, where in the first step we rewrite \cref{model} using \cref{musum} as
\be\label{modelmusum}
\bar{y}_b(\beta) = y_0 \int_0^{E_\text{max}} S(E) D_b(E)\, \exp \Bigl \{-\sum_{k=1}^N \mu_k(E)\beta_k \Bigr \} \,\mathrm{d}E,
\ee
where $\beta=[\beta_1,\dots,\beta_N]$ are the projections given by
\be\label{eq:betak}
\beta_k=\int_L q_k(x)\,dx.
\ee
Inverting the data $y$ gives the `metadata' $\beta$ which are then used as input (for all lines $L$) for the second step of tomographic imaging. Particular expansions of the type in \cref{musum} include representations in terms of the underlying physical processes \cite{Alvarez1976-xv} and in terms of materials. In the latter case, $\mu_k(E)$ are the linear attenuation coefficients of a list of materials obtained from a database, e.g. NIST \cite{nist}, and $q_k(x)$ are the corresponding spatial coefficients. Since the measurements $y$ are described by independent Poisson variables $y_b$, we can write for the likelihood
\be
p(y|\beta ) = \prod_{b=1}^{N_b} p(y_b|\beta),\quad p(y_b|\beta ) = \frac{1}{y_b!} \bar y_b^{y_b} \exp\{ -\bar y_b\}.
\ee
The projections $\beta$ can then be recovered by minimising the Poisson log-likelihood or equivalently by minimising the generalised, discrete Kullback-Leibler (KL) distance, which can be considered as an appropriate data misfit \cite{Hohage2016-da, Hohweiller2017-qd}, given by
\begin{equation}\label{eq:kl}
    D_\text{KL}(y \|\bar{y}) =  y^T \log\left(\frac{y}{\bar{y}} \right) + 1^T (\bar{y} - y),
\end{equation}
where $1$ denotes the $N$-dimensional vector of ones and division is elementwise. The Poisson log-likelihood is sometimes approximated by a weighted Euclidean distance, but it should be noted that this is only a good approximation for high photon counts \cite{Hohage2016-da}. For both the KL and the weighted Euclidean distance, the misfit terms are differentiable, but generally non-convex due to the non-linearity \cite{Kamilis2019-jb, Abascal2018-nl}. This poses a problem to optimisation methods, as they may fail depending on initialisation and may require tuning of algorithmic parameters \cite{Ducros2017-xa}, small step sizes with many iterations \cite{Kamilis2019-jb} or indeed solving additional subproblems \cite{Abascal2018-nl}. A further complexity arises due to the addition of hand-crafted regularisation terms which aim to encode some prior information, such as smoothness or sparsity promoting, on the projections $\beta$. Such an imposition of a general prior is less natural in the projection domain and although the Radon transform is smoothing and a certain level of smoothness in the projections is to be expected for each material and angle, choosing the appropriate regularisation that combines information across materials and angles is not straightforward. Regularisation terms based on mixed norms (as in e.g. Collaborative Total Variation \cite{Duran2016-od}) may be in principle a reasonable choice, but are not equipped to capture the specific features of the task at hand. As we propose in this manuscript, such issues can be circumvented by the use of learning algorithms.

Upon recovery of $\beta_k$ in the first step, the second step of recovering $q_k(x)$ can proceed with modern tomographic reconstruction techniques, although having limited-angle data makes this task challenging. Nevertheless, state-of-the-art methods based on Deep Learning are capable of obtaining images with relatively small error \cite{Arridge2019-wq}, even in the sparse-view setting. Specifically, the learned primal-dual algorithm proposed in \cite{Adler2018-xy} is appealing due to its high reconstruction performance and generality. For this reason, this method will be our main tool for both steps of the imaging process as we describe in the next sections.

Note that inevitably some error will be incurred in the recovery of $\beta$ due to noise effects and algorithm-specific variability. This will essentially be a bias term that may have different magnitude for each material. Therefore, the usual assumption of Gaussian noise for tomographic inversion is generally not valid for the metadata obtained after the first inversion step. If this is not taken into account, it can have a negative effect on the performance of the algorithms.

\subsection{Discrete model}
For the purposes of this manuscript, we restrict our study to the 2D case and use $N_p=N_{p_x}\times N_{p_y}$ pixels as a basis $\{\chi_1(x),\ldots,\chi_{N_p}(x)\}$ for the image domain. Then using a basis of $N$ materials with LACs $\mu_k$, the expansion in \cref{musum} is expressed as
\be\label{musumdiscrete}
\mu(x,E)=\sum_{k=1}^N \mu_k(E) \sum_{i=1}^{N_p} q_{ki} \chi_i(x),
\ee
where the coefficients $q_{ki}\geq 0$ can now be interpreted as the volume fractions of material $k$ in pixel $i$ \cite{Long2014-np}. In particular, when volume conservation is valid, which is a reasonable assumption for the materials found in the human body \cite{Mendonca2014-bk}, the volume fractions obey $q_{ki}\leq 1$ with $\sum_k q_{ki}=1, \; \forall i$. 
\Cref{musumdiscrete} can also be written in terms of Mass Attenuation Coefficients (MACs) $m_k$ and densities $\rho_k$, by using the relation $m_k=\mu_k\rho_k^{-1}$.
Although the typical requirement for material decomposition is $N_b\geq N$, the above constraints can be used to decompose $N+1$ materials using $N_b=N$ energy bins, as well as to constrain and increase the accuracy of the solution by eliminating a linear dependency. A special case is when $q_{ki}$ are restricted to take binary values, which can be used to cast the problem into the discrete tomography setting \cite{Kamilis2019-jb}. We note that these constraints will typically need to be enforced explicitly in the solution of the resulting optimisation problems via e.g. proximal operators \cite{Kamilis2019-jb}, which increases the computational complexity. Contrary to this, data-driven models such as the one proposed here are able to implicitly enforce any constraints by using training data which adhere to them. Effectively, this simplifies the problem and reduces computational costs.

Similarly, we discretise the energy integral using a quadrature rule which gives $N_e$ points $E_j$ within the interval $[0, E_{\text{max}}]$ and corresponding weights $w=[w_1,\dots,w_{N_e}]$, so that \cref{modelmusum} is written as
\be\label{modelmusumdiscrete}
\bar{y}_b(\beta) = y_0 \sum_{j=1}^{N_e} w_j S(E_j) D_b(E_j)\, \exp \Bigl \{ -\sum_{k=1}^N \mu_k(E_j) \beta_k \Bigr \},
\ee
or collecting $S(E_j)$ into a vector $s$, $D_b(E_j)$ into an $N_b\times N_e$ matrix D and $\mu_k(E_j)$ into an $N_e \times N$ matrix $M$, we can write concisely
\be\label{forwdiscrete}
\bar{y}(\beta) = y_0 D \Bigl (w s \exp \bigl \{ - M\beta \bigr \} \Bigr ),
\ee
where multiplication of vectors is elementwise. For numerical stability, we compute \cref{forwdiscrete} in the $\log$ domain and use a stable implementation of the $\texttt{LogSumExp}$ operation for computing energy domain sums. For later use, we also require the adjoint of the derivative with respect to $\beta$, given by
\be\label{eq:deriv}
\partial\overline{y}^*(\beta)=-y_0 M^T \,\text{diag}\Bigl (w s \exp \bigl \{-M\beta \bigr \} \Bigr ) D^T.
\ee

Up to this point, our expressions refer to a single arbitrary line (X-ray trajectory) $L$. We now assume $N_\theta$ number of projection angles and $N_d$ detector elements for a total $N_\theta\times N_d$ rays. All preceding equations are then extended trivially by applying operations pointwise for each ray $L(\theta, l)$ that corresponds to angle $\theta$ and detector element $l$. For the spectral mixing, we define the non-linear operator $\bar{y}(\beta): Y\rightarrow Z$, with $Y=\mathbb{R}^{N\times N_\theta\times N_d}$ and $Z=\mathbb{R}^{N_b\times N_\theta \times N_d}$. For the projections, we define the linear operator $R(q)=\sum_{i=1}^{N_p} w_{i \theta l} q_{ki}: X\rightarrow Y$, with $X=\mathbb{R}^{N \times N_p}$ and $w_{i \theta l}=\int_{L(\theta,l)} \chi_i(x) \,dx$ being the contribution of each pixel $i$ to the line integral for the line $L(\theta,l)$.

\subsection{Classical optimisation}

\begin{algorithm}[h]
		 \caption{General classical primal-dual algorithm}\label{alg:primaldual}
		\begin{algorithmic}[1]
		\State \textbf{Input:} $\mathcal{D},A,\lambda\mathcal{R},\mathcal{C},d,\tau$
		\State \textbf{Initialization:} $u^0, z^0, k=0$
			\While{convergence criterion is not met}
				\vspace{0.1cm}
				\State $z^{k+1}=\Gamma^d_{\mathcal{D},\lambda\mathcal{R},\mathcal{C}}(z^k, \tau, A(u^{k}), d)$
				\State $u^{k+1}=\Gamma^p_{\mathcal{D},\lambda\mathcal{R},\mathcal{C}}(u^k, \tau, \partial A^*(u^k)(z^{k+1}))$
			    \State $k=k+1$
				\vspace{0.1cm}
			\EndWhile
			\State \Return $u^{k}$
	\end{algorithmic}
	\end{algorithm}

The two subproblems can be formulated as optimisation problems for a general functional $\mathcal{F}$
\be\label{eq:classoptim}
\mathcal{F}(u)=\mathcal{D}(A(u); d)+\lambda\mathcal{R}(u) + \mathcal{C}(u), 
\ee
where $\mathcal{D}$ corresponds to the misfit for data $d$ and forward operator $A$, $\mathcal{R}$ is the regularisation term with regularisation parameter $\lambda$, and $\mathcal{C}$ encodes constraints. For the unmixing problem with $u=\beta$, $d=y$ and $A(\beta)=\bar{y}(\beta)$, suitable choices are the KL distance as in \cref{eq:kl}, a Collaborative Total Variation regularisation term $\mathcal{R}$ \cite{Duran2016-od} and an indicator functional $\mathcal{C}$ for the simplex $\Delta$ given by
\be\label{eq:simplex}
\Delta \doteq \left\{ \beta\in Y | \sum_{k=1}^N \beta_{k\theta l} = |L(\theta,l)|\, \text{and}\, \beta\geq 0\right\},
\ee
as can be derived from volume conservation. Similarly, for the $N$ tomography problems, suitable choices with $u=q$, $d=\beta$ and $A(q)=R(q)$, are the Euclidean distance as data misfit, a Total Variation regularisation term and a positivity constraint. 

The tomography problem can be solved using convex optimisation methods \cite{Benning2018-ow} such as the Alternating Direction Method of Multipliers (ADMM) or a primal-dual method in general \cite{Parikh2014-wq}, using an appropriate formulation of \cref{eq:classoptim} and the required proximal operators. The unmixing problem poses a greater challenge as it is non-linear and generally non-convex, which may cause convex optimisation methods to fail to converge to a global minimum if the initial guess is not optimal. Nevertheless, a variant of ADMM with non-linear operator constraint \cite{Benning2018-ow} has shown in practice very good performance for a case of 3 materials \cite{Kamilis2019-jb}, albeit with high computational costs. Formulations based on the Bregman distance provide a more robust alternative, but require the solution of subproblems which can increase the computational cost even further \cite{Ducros2017-xa}. Moreover, for both methods, the performance generally deteriorates significantly with the addition of more materials as the problem becomes more ill-posed in that case.

As shown in \cref{alg:primaldual}, both the linearised ADMM and the ADMM with a non-linear operator constraint can be formulated as iterative updates of the primal variable $u$ and a dual variable $z$ using operators $\Gamma^p$ and $\Gamma^d$ that are in general based on suitable expressions of the proximal operators for $\mathcal{D}$, $\mathcal{R}$, $\mathcal{C}$, acting on primal and dual variables via the operator $A$ and the adjoint of its Fr\'{e}chet derivative $\partial A^*$ (in practice, $A$ may be replaced by a stacked operator to allow an efficient formulation). Apart from the regularisation parameter $\lambda$, the algorithms also require a user-chosen step size $\tau$ for the primal and/or dual updates, which can be cumbersome to fine-tune for a specific problem. Even more generally, the choice of $\mathcal{D}, \mathcal{C}$ and particularly $\mathcal{R}$ may not be optimal for a specific application. As we describe in the next section, data-based methods can provide a competitive alternative to overcome such problems.

\section{Learned Spectral CT}

The starting point is to replace the proximal operators $\Gamma^p$ and $\Gamma^d$ with general operators $\Gamma^p_{\omega^p}$ and $\Gamma^d_{\omega^d}$ that are parameterised by a sequence of parameters $\omega^p$ and $\omega^d$ respectively. Effectively, these operators generalise $\mathcal{D}$, $\lambda\mathcal{R}$, and $\mathcal{C}$ to implicitly learn a suitable data-fit term, a prior information term and any constraints, as well as an implicit step size $\tau$. The physics of the model are still imparted through the operator $A$, therefore increasing the stability and effectiveness of the learned algorithm even with moderately-sized training data. Additionally, more flexibility is inserted into the algorithm by i) allowing the weights to change at each iteration, hence denoted as $\omega^p_k$ and $\omega^d_k$ for the $k$-th iteration and ii) extending the primal variable as $u=[u_1, \dots, u_{n_\text{primal}}]$ and dual variable as $z=[z_1,\dots,z_{n_\text{dual}}]$ to have memory between iterations of size $n_{\text{primal}}\geq 2$ and $n_{\text{dual}}\geq 1$ respectively.  The general form of the algorithm is outlined in \cref{alg:learnedalgo}. Note that an important difference of the classical and learned algorithms is that the first is typically run for many iterations and has theoretical convergence guarantees, while the second is run for a small, fixed number of iterations due to computational cost considerations (in the training phase) and lack of convergence guarantees.

Given a training dataset of $n_\text{train}$ pairs $\{u, d\}$, the training phase of the learned primal-dual algorithm with $n_\text{{iter}}$ iterations involves the tuning of $\omega^p, \omega^d$, where the objective is to minimise an empirical loss function $\mathcal{L}$ as
\be\label{eq:mse}
\hat{\omega}^p,\hat{\omega}^d\in \argmin  \mathcal{L}(u,T_{n_{\text{iter}}}(d)),
\ee
where $T_{n_{\text{iter}}}(d)$ denotes the output of the algorithm for given $\omega^p,\omega^d$ and data $d$.

What remains is to choose representations of the operators $\Gamma^p_{\omega^p_k}$ and $\Gamma^d_{\omega^d_k}$ such that i) they are complex enough to capture the sought features, ii) they can be applied effectively once trained, and iii) the problem in \cref{eq:mse} can be solved efficiently via e.g. first-order methods. Such requirements are met by Deep Neural Networks (DNNs), so that the operators are represented by one or more layers of the network and the parameters $\omega^p_k$ and $\omega^d_k$ encode the weights and biases. In addition, the operators $A$ and $\partial A^*$ are embedded as fixed layers within the networks. We proceed to describe the chosen DNN architecture and the two ways that the learned algorithms for unmixing and imaging can be arranged and trained, namely as separate networks or integrated with end-to-end training.

\subsection{DNN architecture}
We represent the operators as in \cite{Adler2018-xy} using residual networks composed of convolutional layers and Parametric Rectified Linear Unit (PReLU) activation functions. The choice of convolutional layers is due to their computational effectiveness and favourable properties such as translation invariance. For the $k$-th iteration, the operator $\Gamma^p_{\omega_p^k}$ is represented as
\be\label{eq:primalopnet}
\Gamma^p_{\omega^p_k} = I + C^p_{k_3}\circ \sigma^p_{k_2} \circ C^p_{k_2} \circ \sigma^p_{k_1} \circ C^p_{k_1},
\ee
where $I$ is the identity operator, $C^p_{k_i}$ the $i$-th convolutional layer with weights $w^p_{k_i}$ and biases $b^p_{k_i}$, and $\sigma^p_{k_i}$ the $i$-th PReLU activation function given as
\be
\sigma^p_{k_i}(x)= \begin{cases} x & x\geq 0 \\
-c^p_{k_i} & x<0, \end{cases}
\ee
with weights $c^p_{k_i}$. We can then write $\omega^p$ as the collection of all the weights in all iterations, given as
\be
\omega^p= \bigl \{\omega^p_1,\dots,\omega^p_{n_\text{iter}} \bigr \},
\ee
with each $\omega^p_k$ given by the collection of the corresponding weights and biases as
\be
\omega^p_k = \bigl \{c^p_{k_1},c^p_{k_2},w^p_{k_1},w^p_{k_2},w^p_{k_3},b^p_{k_1},b^p_{k_2},b^p_{k_3} \bigr \}.
\ee
A similar representation is used for the dual operator $\Gamma^d_{\omega_k^d}$.

In the following we denote by $T^\text{unmix}_{n_{\text{unmix}}}$ with $n_{\text{unmix}}$ iterations, the learned algorithm corresponding to the unmixing step and by $T^{\text{reco}}_{n_\text{reco}}$ with $n_\text{reco}$ iterations the learned algorithm corresponding to the imaging step. The two parts of the inverse problem differ in the dimensions of the input and output variables and therefore the implementation details of the neural networks are also different. Specifically, for the imaging task the primal operator in \cref{eq:primalopnet} has input dimensions $N_{p_x}\times N_{p_y}\times (n_\text{primal}+1)$ where the last dimension is obtained from the concatenation of $u=q$ (assuming memory $n_\text{primal}$)  with $\partial A^*(u_1^k)(z_1^{k+1})$. The dual operator has input dimensions $N_\theta\times N_d\times (n_\text{dual}+2)$ where as before the last dimension is due to concatenation. The convolutional layers in this case are 2D, with the last dimension of the input operating as the channel dimension. On the other hand, for the learned unmixing task, the primal operator has input dimensions $N\times N_\theta\times N_d\times (n_\text{primal}+1)$ and the dual operator has input dimensions $N_b \times N_\theta \times N_d\times (n_\text{dual}+2)$. The convolutional layers can also be 2D in this case but to allow the network to learn correlations between materials and bins, it is also possible to use 3D convolutions that operate across the first three dimensions. We have found that this choice gives better results in some cases, see for example the comparison for random-ellipse phantoms in the results section. However, due to memory limitations we were unable to use 3D convolutions for examples with smaller pixel size.
\subsection{Training methods}
\subsubsection{Separate learned unmixing and learned reconstruction (SL)}
In the simplest case, the two learned primal-dual algorithms can be arranged and trained separately. For learned unmixing the training objective is to optimise $\omega^p, \omega^d$ using pairs of data $\{\beta, y\}$, so that $\mathcal{L}(\beta, \hat{\beta})$ is minimised with $\hat{\beta}=T^{\text{unmix}}_{n_\text{unmix}}(y)$. For learned reconstruction, the objective is to minimise $\mathcal{L}(q, \hat{q})$ using pairs of data $\{q, \beta\}$, with $\hat{q}=T^{\text{reco}}_{n_\text{reco}}({\beta})$. In the second step, the $\beta$ used in training can be simulated from the phantoms $q$ and corrupted with e.g. Gaussian noise. However, this ignores any systematic errors in the unmixing step. Therefore, a better alternative is to first train the unmixing algorithm and then use the recovered $\hat{\beta}$. This allows the imaging algorithm to learn and correct any errors produced in the first step. Since the two steps are trained separately, there is still a limited flexibility in the training process to optimise for the final task of imaging. The next approach is better suited to this problem.

\subsubsection{Integrated learned unmixing and reconstruction (IL)}
\begin{figure*}[!t]
    \centering
    \includegraphics[width=0.67\textwidth]{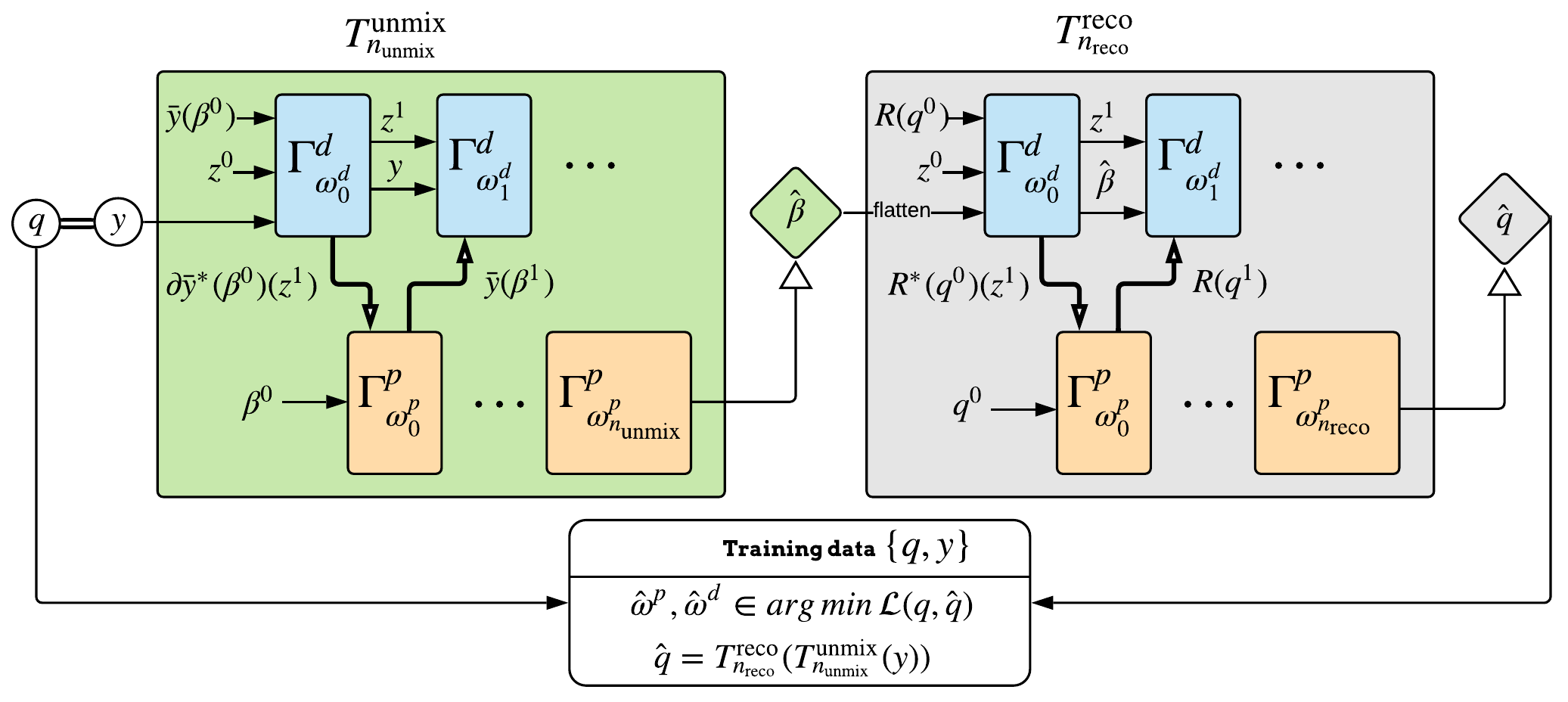}%
    \caption{Illustration of the proposed integrated learned unmixing and imaging method (IL).}
    \label{fig:network}
\end{figure*}

In this approach, the two networks are integrated and trained end-to-end, so that given data $\{q, y\}$, the objective is to minimise $\mathcal{L}(q, \hat{q})$ where $\hat{q}=T^{\text{reco}}_{n_\text{reco}}(T^{\text{unmix}}_{n_\text{unmix}}(y))$. Note that the composition of the two networks is possible by adding a connecting flattening layer across the material dimension from the output of the final primal variable of the unmixing step to the input of the first dual iterate of the reconstruction step. This method ensures that the two learned networks are optimised for the final task of imaging and automatically takes into account any error induced by the first unmixing step. As we report in the results section, this approach performs better than SL and is our method of choice for learned Spectral CT. An illustration of the approach is shown in \cref{fig:network}.

\begin{algorithm}[h]
	\caption{Learned primal-dual algorithm}\label{alg:learnedalgo}
	\begin{algorithmic}[1]
		\State \textbf{Input:} $A,d$
		\State \textbf{Initialisation:} $u^0, z^0, k=0$
			\While{$k<n_{\text{iter}}$}
				\vspace{0.1cm}
				\State $z^{k+1}=\Gamma^d_{\omega_k^d}([z^k, A(u_2^{k}), d])$
				\State $u^{k+1}=\Gamma^p_{\omega_k^p}([u^k, \partial A^*(u_1^k)(z_1^{k+1})])$
			    \State $k=k+1$
				\vspace{0.1cm}
			\EndWhile
			\State \Return $u^{k}_1$
	\end{algorithmic}
	\end{algorithm}

\section{Results}
\begin{figure}[t!]
    \centering
    \includegraphics[width=0.5\textwidth]{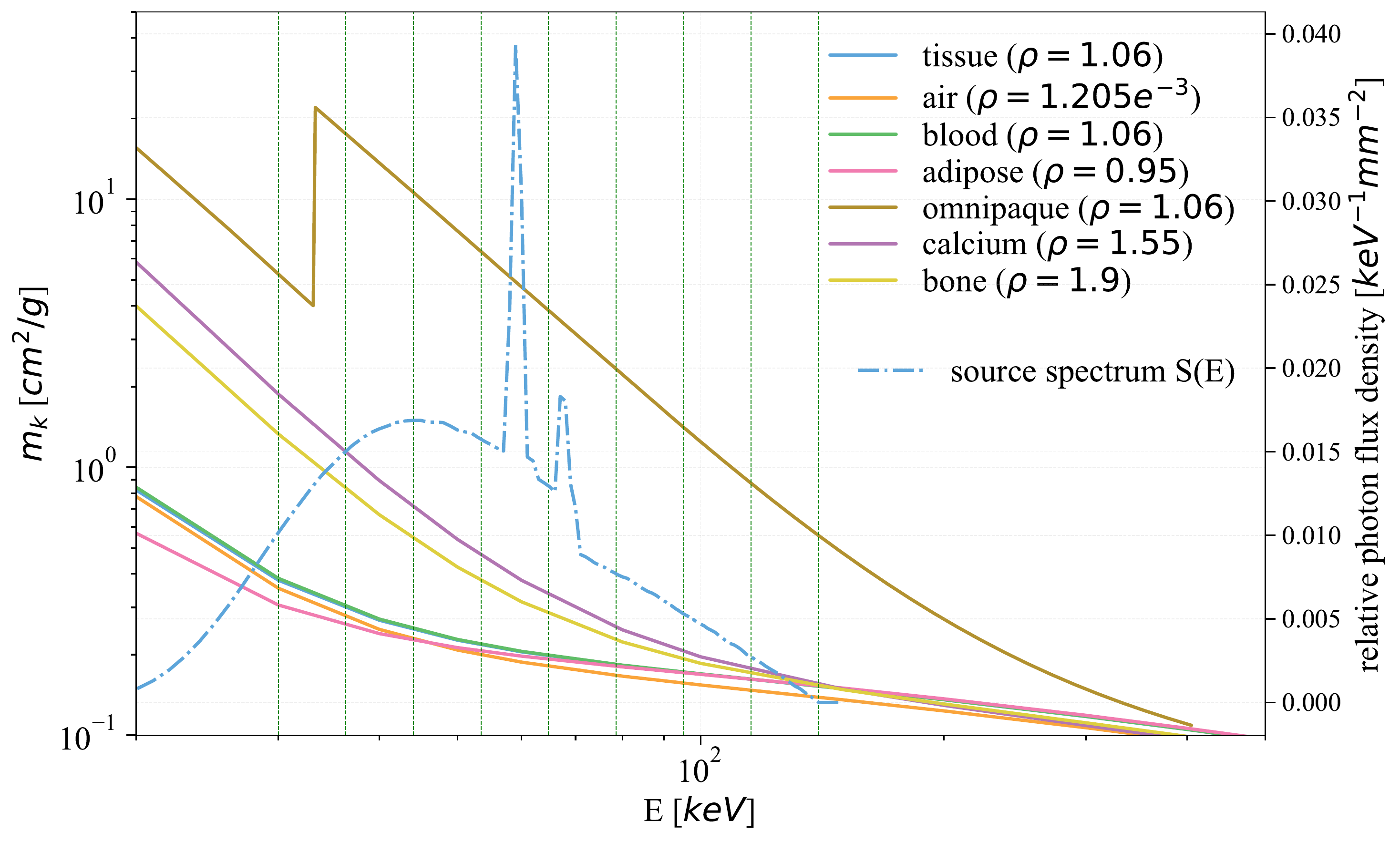}
    \caption{Material MACs $m_k(E)$ and normalised source spectrum $S(E)$ used in the numerical examples. Vertical dashed lines represent the $N_b=8$ energy bins used between $E_{\text{min}}=$ \SI{30}{keV} and $E_{\text{max}}=$ \SI{140}{keV}. Material densities $\rho_k$ in \si{g/cm^3}. Data taken from NIST \cite{nist}.}
    \label{fig:macs}
\end{figure}

We present numerical results for synthetic datasets generated from i) a random-ellipse phantom which serves as an example of data that have a high degree of variation, and ii) a medical phantom of the heart area that has more structure in the data and serves as a more realistic benchmark. In both cases we assume a material basis with $N=5$ consisting of bone, tissue, calcium, air and with the fifth material varying between adipose tissue, blood or omnipaque. \Cref{fig:macs} shows the material MACs $m_k(E)$, together with the source spectrum $S(E)$ of a \SI{140}{kVp} source and the $N_b=8$ energy bins which are equally distributed in log-space between $E_\text{min}=$ \SI{30}{keV} and $E_\text{max}=$ \SI{140}{keV}. The energy integral was approximated using a Fej\'{e}r quadrature rule within the same range with $N_e=16$. The discretisation details for the chosen datasets are shown in \cref{tab:dataparams}. For all cases we set $n_{\text{unmix}}=n_{\text{reco}}=10$.

\begin{table}[t!]
    \centering
        \caption{Dataset parameters}\label{tab:dataparams}
    \begin{tabular}{p{.8cm}|p{1.1cm}|p{.4cm}|p{.4cm}|p{.7cm}|p{.4cm}|p{.9cm}|p{.5cm}}
        Dataset & Type & $N_{p_x}$ & $N_{p_y}$ & Pixel size [\si{\cm}] & $N_d$ & Detector el. size [\si{\cm}]& $y_0$ \\
        \hline
        $\texttt{e\_5}$ & ellipses & $128$ & $128$ & $1.0$ & $183$ & $1.0$ & $10^{12}$ \\
        $\texttt{m\_5}$ & medical & $512$ & $512$ & $0.125$ & $727$ & $0.125$ & $10^9$ \\
        $\texttt{m\_5\_p}$ & medical $+$ plaque & $512$ & $512$ & $0.125$ & $727$ & $0.125$ & $10^9$%
    \end{tabular}
\end{table}
\subsubsection{Random-ellipse phantom (dataset $\texttt{e\_5}$)}
This type of phantom contains a set of ellipses drawn from a Poisson distribution with mean $N_\text{ellipses}=25$. The ellipses vary randomly in size, shape,
position, orientation and are randomly assigned a material from the chosen list of materials (except air). Then the remaining pixels in the domain are chosen as air. For validation purposes, an altered version of the Shepp-Logan phantom was also used.
\subsubsection{Medical phantom (dataset $\texttt{m\_5}$)}
All phantom data for the medical use-case were generated using the 4D XCAT phantom \cite{Segars2010-wa}. The program is based on real CT data and allows for generating a wide range of anatomical features. All slices in the datasets were chosen to be located in the upper torso area to include the heart. The training and test data were generated by varying the length, width, and height, as well as by changing the spatial orientation of the phantoms, which were also randomly altered between male and female. The materials were assigned to the phantoms by selecting regions with certain attenuation coefficients from the XCAT output. To make the phantoms more realistic, overlapping materials were allowed with two materials at $50\%$ volume fraction each.
\subsubsection{Medical phantom with plaque (dataset  $\texttt{m\_5\_p}$)}
To go towards the use-case of atherosclerosis, the phantoms were modified
to contain a single small patch of plaque. This was
generated in the left anterior descending vessel and scaled by
five iterations of binary dilation. We also used Omnipaque for this experiment.
\subsubsection*{Implementation details}
The implementation of the operators $A$ and $\partial {A}^*$ was done using Python, the ODL library \cite{odl} and the ASTRA toolbox \cite{VanAarle2015} making use of GPU acceleration. Reference classical optimisation methods were implemented in ODL, while the NNs were implemented using Tensorflow 1.14 \cite{Abadi2016-na} with standard layers and by additionally adding the operators $A$, $\partial A^*$ as custom Tensorflow layers.

The convolutional layers were 2D except for an experiment for dataset $\texttt{e\_5}$ and in all cases they had 32 filters with filter size $3\times 3$. Xavier initialisation was used for the NN weights $w$ and the biases $b$ were initially set to zero. We further used the ADAM optimiser for training, with default parameters, except $\beta_2 = 0.99$. For the learning rate,
cosine annealing was used with initial value $10^{-3}$. In addition, we applied
global gradient norm clipping, limiting the gradient’s norm to $1$. The batch size varied between the experiments, with batch size 1 enforced for the medical datasets as anything larger caused memory issues. As loss function $\mathcal{L}$ we used the mean square error.

For evaluation, a test dataset with the same parameters but different random
seed was used. Results are reported with metrics: i) structural similarity index (SSIM), normalised
root-mean-square error (NRMSE) and peak signal-to-noise-ratio (PSNR) for 100 samples and averaged per material. For the random-ellipse dataset, the Shepp-Logan phantom was additionally evaluated.

We performed extensive numerical tests with different optimisation and regularisation methods and we chose as a reference method the ADMM with non-linear operator constraint and simplex constraint \cref{eq:simplex}, and the linearised ADMM with TV regularisation for imaging. 

\begin{figure}[!t]
    \centering
    \includegraphics[width=0.225\textwidth]{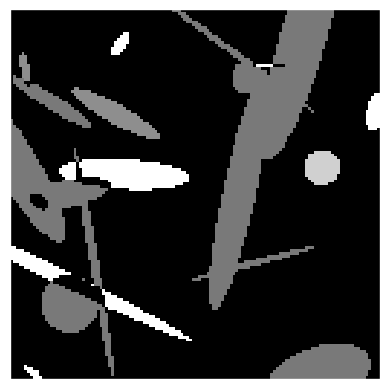}%
    \includegraphics[width=0.225\textwidth]{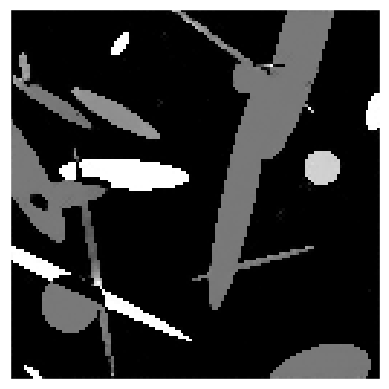}%
    \includegraphics[width=0.0370\textwidth]{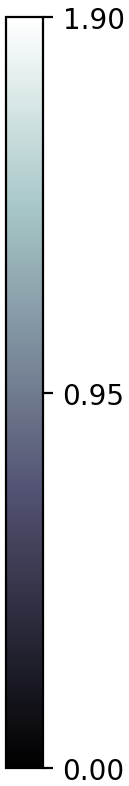}
    \caption{Comparison of densities (grey levels) for a ground truth random-ellipse phantom (left) with 5 materials (dataset $\texttt{e\_5}$) including adipose tissue and the corresponding reconstruction (right) given by IL method (see also \cref{tab:integratedalgobest}). Densities within \numrange{0}{1.9} \si{g/cm^{-3}}.}
    \label{fig:adipose_combined}
\end{figure}

\begin{figure*}[!t]
    \centering
    \includegraphics[width=0.40\textwidth]{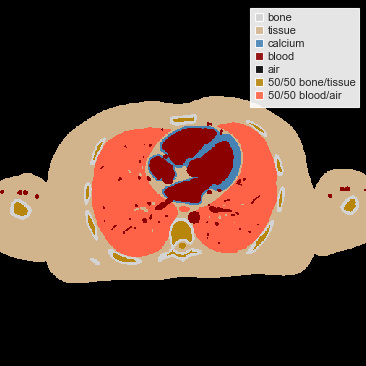}%
    \hspace{0.3cm}
    \includegraphics[width=0.40\textwidth]{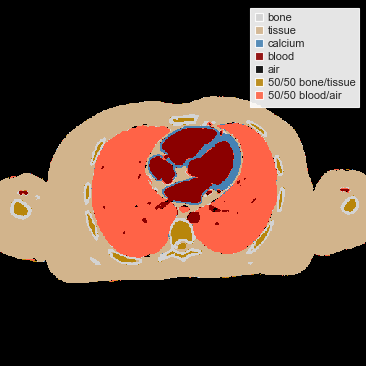}%
    \caption{Medical phantom from dataset $\texttt{m\_5}$ with 5 materials including blood and mixed material regions. Left: Ground truth. Right: Reconstructed phantom (see also \cref{tab:medical_phantom}).}
    \label{fig:full_example}
\end{figure*}

\begin{figure*}[!t]
    \centering
    \includegraphics[width=0.40\textwidth]{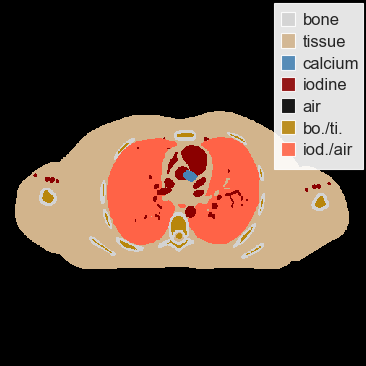}%
    \hspace{0.3cm}
    \includegraphics[width=0.40\textwidth]{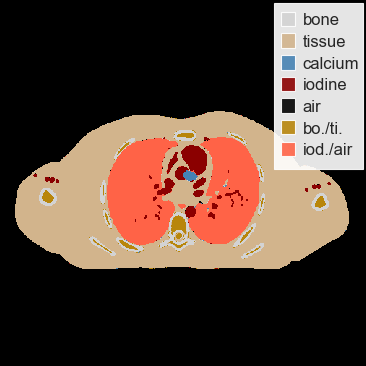}
    \caption{Medical phantom from dataset $\texttt{m\_5\_p}$ using Omnipaque and with a single, enlarged plaque in the left anterior descending vessel. Left: Ground truth. Right: Reconstructed phantom (see also \cref{tab:medical_phantom_plaque}).}
    \label{fig:large_plaque}
\end{figure*}

The results for the random ellipse-phantom using SL are shown in \cref{tab:separatealgo}, while \cref{tab:integratedalgocompare} shows the same case using IL (with 2D convolutions), which clearly outperforms the first method with high SSIM. Even better results with IL are shown in \cref{tab:integratedalgobest} using this time 3D convolutional layers and an increased number of training iterations. An image reconstruction example with this last setup is shown in \cref{fig:adipose_combined}. For the medical phantom with fifth material being blood, \cref{tab:medical_phantom} shows results using IL, in comparison with the reference optimisation approach. Clearly, there is a substantial improvement in both SSIM and NRMSE with the proposed learned Spectral CT method. Additionally, on our workstation, it took \SI{0.23}{s} to produce an image with the learned approach (once trained which required \SI{19}{h}), compared to \SI{37}{s} with the reference method (parallelised). An example of phantom reconstruction in this case is depicted in \cref{fig:full_example}, which shows good identification of the material regions with only some small features misidentified. Results for the case of a medical phantom with an added enlarged plaque region are shown in \cref{tab:medical_phantom_plaque} with high SSIM, and a corresponding image reconstruction is shown in \cref{fig:large_plaque} which shows that the calcified plaque region is clearly identified.

\begin{table}[!t]
\centering
    \renewcommand{\arraystretch}{1.3}
    \begin{threeparttable}
	\caption{SSIM, NRMSE and PSNR of random-ellipse (top) and Shepp-Logan (bottom) material phantoms with SL\tnote{1}.}
	\label{tab:separatealgo}
    \begin{tabular}{l c c c c c c c}
        \hline
        & \textbf{Bone} & \textbf{Tissue} & \textbf{Calc}. & \textbf{Adip.} & \textbf{Air} & avg.\tnote{3}\\
        \hline
        SSIM & 0.773 & 0.721 & 0.786 & 0.653 & 0.716 & \textbf{0.730}\\
        NRMSE & 0.674 & 0.682 & 0.704 & 0.791 & 0.264 & \textbf{0.623} \\
        PSNR\; (\si{dB}) & 17.98 & 16.87 & 19.07 & 16.13 & 13.84 & \textbf{16.78}\\
        \hline
        SSIM\tnote{2} & 0.756 & 0.721 & 0.798 & 0.780 & 0.889 & \textbf{0.789}\\
        NRMSE\tnote{2} & 0.810 & 0.511 & 2.122 & 1.013 & 0.202 & \textbf{0.932}\\
        PSNR\tnote{2}\; (\si{dB}) & 15.50 & 10.17 & 20.42 & 11.00 & 16.83 & \textbf{14.78}\\
        \hline
    \end{tabular}
    \begin{tablenotes}
    \item[1] Unmixing network trained on dataset $\texttt{e\_5}$ with adipose tissue for 20000 iterations and batch size 3. Imaging network trained for 20000 iterations with batch size 15.
    \item[2] Shepp-Logan phantom.
    \item[3] Average values over all materials.
    \end{tablenotes}
    \end{threeparttable}
\end{table}

\begin{table}[!t]
\centering
    \renewcommand{\arraystretch}{1.3}
    \begin{threeparttable}
	\caption{SSIM, NRMSE and PSNR of random-ellipse phantoms (top) and Shepp-Logan (bottom) material phantoms with IL\tnote{1}.}
	\label{tab:integratedalgocompare}
    \begin{tabular}{l c c c c c c}
        \hline
        & \textbf{Bone} & \textbf{Tissue} & \textbf{Calc}. & \textbf{Adip.} & \textbf{Air} & avg.\tnote{3}\\
        \hline
        SSIM & 0.933 & 0.844 & 0.939 & 0.805 & 0.959 & \textbf{0.896}  \\
        NRMSE & 0.270 & 0.502 & 0.391 & 0.954 & 0.092 & \textbf{0.442} \\
        PSNR\; (\si{dB}) & 26.46 & 20.55 & 26.23 & 18.40 & 23.34 & \textbf{23.00} \\
        \hline
        SSIM\tnote{2} & 0.785 & 0.738 & 0.901 & 0.790 & 0.994 & \textbf{0.842} \\
        NRMSE\tnote{2} & 0.982 & 0.515 & 0.970 & 1.011 & 0.020 & \textbf{0.495} \\
        PSNR\tnote{2}\; (\si{dB}) & 13.83 & 10.10 & 27.22 & 11.01 & 37.01 & \textbf{20.23} \\
        \hline
    \end{tabular}
    \begin{tablenotes}
    \item[1] Trained on dataset $\texttt{e\_5}$ with adipose tissue for 20000 iterations at batch size 3. 
    \item[2] Shepp-Logan phantom.
    \item[3] Avearage values over all materials.
    \end{tablenotes}
    \end{threeparttable}
\end{table}

\begin{table}[!t]
\centering
    \renewcommand{\arraystretch}{1.3}
    \begin{threeparttable}
	\caption{SSIM, NRMSE and PSNR of random-ellipse (top) and Shepp-Logan (bottom) material phantoms with IL.\tnote{1}.}
	\label{tab:integratedalgobest}
    \begin{tabular}{l c c c c c c}
        \hline
        & \textbf{Bone} & \textbf{Tissue} & \textbf{Calc}. & \textbf{Adip.} & \textbf{Air} & avg.\tnote{3}\\
        \hline
        SSIM & 0.983 & 0.965 & 0.985 & 0.925 & 0.990 & \textbf{0.970}\\
        NRMSE & 0.104 & 0.202 & 0.144 & 0.391 & 0.045 & \textbf{0.169} \\
        PSNR\; (\si{dB}) & 34.36 & 29.66 & 35.46 & 24.64 & 30.27 & \textbf{30.88}\\
        \hline
        SSIM\tnote{2} & 0.849 & 0.726 & 0.925 & 0.591 & 0.997 & \textbf{0.818}\\
        NRMSE\tnote{2} & 0.595 & 0.532 & 0.616 & 0.988 & 0.007 & \textbf{0.548}\\
        PSNR\tnote{2}\; (\si{dB}) & 18.18 & 9.82 & 31.17 & 11.21 & 45.72 & \textbf{23.22}\\
        \hline
    \end{tabular}
    \begin{tablenotes}
    \item[1] Trained on dataset $\texttt{e\_5}$ with adipose tissue for 30000 iterations at batch size 10, using 3D convolutions for the unmixing.
    \item[2] Shepp-Logan phantom.
    \item[3] Average values over all materials.
    \end{tablenotes}
    \end{threeparttable}
\end{table}

\begin{table}[!t]
\centering
    \renewcommand{\arraystretch}{1.3}
    \begin{threeparttable}
	\caption{SSIM, NRMSE and PSNR of medical phantoms with 5 materials including blood, with IL\tnote{1}.}
	\label{tab:medical_phantom}
    \begin{tabular}{l c c c c c c}
        \hline
        & \textbf{Bone} & \textbf{Tissue} & \textbf{Calc}. & \textbf{Adip.} & \textbf{Air} & avg.\tnote{3}\\
        \hline
        SSIM & 0.998 & 0.978 & 0.999 & 0.980 & 0.993 & \textbf{0.990}\\
		NRMSE & 0.119 & 0.114 & 0.107 & 0.174 & 0.023 & \textbf{0.107} \\
		PSNR\; (\si{dB}) & 36.35 & 26.46 & 41.82 & 28.73 & 34.71 & \textbf{33.65}\\
        \hline
        \hline
        SSIM classical\tnote{2} & 0.528 & 0.258 & 0.895 & 0.228 & 0.549 & \textbf{0.492}\\
        NRMSE classical\tnote{2} & 0.928 & 0.761 & 0.999 & 0.817 & 0.354 & \textbf{0.772}\\
        \hline
    \end{tabular}
    \begin{tablenotes}
    \item[1] Trained on dataset $\texttt{m\_5}$ for 15000 iterations at batch size 1.
    \item[3] Reference classical optimisation method.
    \item[4] Average values over all materials.
    \end{tablenotes}
    \end{threeparttable}
\end{table}

\begin{table}[!t]
\centering
    \renewcommand{\arraystretch}{1.3}
    \begin{threeparttable}
	\caption{SSIM, NRMSE, PSNR of medical phantoms with 5 materials including Omnipaque and calcified plaque, with IL\tnote{1}.}
	\label{tab:medical_phantom_plaque}
    \begin{tabular}{l c c c c c c}
        \hline
        & \textbf{Bone} & \textbf{Tissue} & \textbf{Calc}. & \textbf{Omnip.} & \textbf{Air} & avg.\tnote{2}\\
        \hline
        SSIM & 0.996 & 0.991 & 1.000 & 0.995 & 0.994 & \textbf{0.995}\\
        NRMSE & 0.195 & 0.079 & 0.241 & 0.100 & 0.021 & \textbf{0.127}\\
        PSNR\; (\si{dB}) & 32.28 & 29.14 & 43.56 & 35.09 & 35.12 & \textbf{35.04}\\
        \hline
    \end{tabular}
    \begin{tablenotes}
    \item[1] Trained on dataset $\texttt{m\_5\_p}$ for 15000 training iterations at batch size 1.
    \item[2] Average values over all materials.
    \end{tablenotes}
    \end{threeparttable}
\end{table}

\section{Conclusion}
We have proposed a data-driven, model-informed approach for fast SPCCT material identification and imaging, based on two learned primal-dual algorithms for each of the steps of unmixing and imaging. We have shown with numerical examples, that this approach correctly identifies most material regions with high SSIM in all cases. In the medical use-case of interest, it outperforms classical optimisation approaches in both SSIM and NRMSE, while being also significantly faster. It correctly images a medical phantom example with a calcified plaque showing the potential of the method for estimating CHD risk.

\section*{Acknowledgment}

The authors would like to thank Michelle C. Williams of the Centre for Cardiovascular Science, University of Edinburgh for providing valuable feedback on the medical case study.

\ifCLASSOPTIONcaptionsoff
  \newpage
\fi

\bibliographystyle{IEEEtran}
\bibliography{IEEEabrv,biblio}

\end{document}